\documentclass[aps,pre,twocolumn,groupedaddress,showpacs]{revtex4-2}

\usepackage{graphicx}
\usepackage{color}
\usepackage{amsmath}

\def\be{\begin{equation}}
\def\en{\end{equation}}
\def\bea{\begin{eqnarray}}
\def\ena{\end{eqnarray}}

\newcommand{\av}[1]{\langle{#1}\rangle}

\begin{document}

\title{Contribution of internal degree of freedom of 
soft molecules to Soret effect}

\author{Takeaki Araki and Natsumi Chikakiyo}

\affiliation{
Department of Physics, Kyoto University, Kyoto 606-8502, Japan\\
}

\date{\today}

\begin{abstract}
We studied the Soret effect in binary dimer-monomer mixtures using non-equilibrium molecular dynamics simulations and investigated the pure contribution of the internal degree 
of freedom of flexible molecules to the Soret effect. We observed that the thermal diffusion factor tends to decrease and change its sign as the molecules become softer. We proposed two possible mechanisms of our observations: change of the molecule structures with the temperature, causing bulkier molecules to migrate to the hotter region; asymmetry of the restitution between rigid and flexible molecules, due to which flexible molecules show larger restitution when placed at the hotter region.

\end{abstract}

\maketitle

\section{introduction}

When a fluid mixture is subjected to a temperature gradient, 
concentration gradients build up. 
This phenomenon is the Soret effect or thermal diffusion
\cite{
deGroot_book_1985,
Wiegand_JPCM_2004,
Platten_JAM_2005,
Srinivasan_book_2012}, 
which is widely observed 
in mixtures of small molecules
\cite{Kita_JCP_2004,Perronace_JCP_2002}, 
polymers \cite{
Giddings_Macro_1976,
Schimpf_JPCB_1989,
Zhang_JCP_2006,
Wuerger_PRL_2007,
Stadelmaier_Macro_2009,
Wuerger_PRL_2009}, 
colloids \cite{
Piazza_SM_2008,
Wuerger_PRL_2007}, 
and biomolecules 
\cite{Duhr_PNAS_2006,
Wienken_NatComm_2010}. 
Since the 19\,th century, 
many experimental, theoretical and numerical studies have been 
reported \cite{
Eastman_JACS_1928,
Artola_JACS_2008,
Artola_MP_2013,
Hafskjold_MP_1993,
Kempers_JCP_1989,
Garriga_JSP_2002,
Wuerger_CRS_2013,
Galliero_FPE_2003,
Debuschewitz_PRL_2001,
Reith_JCP_2000,
VillainGuillot_PRE_2011}. 
However, its physical mechanisms are still not understood completely. 

The Soret effect is caused by some factors; 
its origin is often considered the 
equilibrium thermodynamics \cite{Eastman_JACS_1928,Wuerger_CRS_2013}.
If the chemical potential of each species depends on the temperature, 
the concentration gradient is induced under the temperature gradient 
to homogenize this potential. 
The isotope effect, in which the mass difference 
causes the concentration gradient, 
is difficult to explain 
based on the equilibrium thermodynamics \cite{
Artola_JACS_2008,
Artola_MP_2013,
Galliero_FPE_2003,
Debuschewitz_PRL_2001,
Reith_JCP_2000,
VillainGuillot_PRE_2011}. 
The heavier molecules tend to migrate to the cold side, while 
the lighter ones move to the hot side. 
It is also known that 
differences in diameter, moment of inertia, interaction 
between the solute and the solvent, and/or those among the solutes 
contribute significantly in the Soret effect 
\cite{Wiegand_JPCM_2004}. 

When considering the Soret effect, the molecules are 
usually treated as rigid objects, while the internal degrees of 
freedom of the molecules are ignored. 
If flexible, 
the molecules are thermalized and adapt their structures 
to the local environment (the temperature, the pressure, 
and the concentration). 
These changes may influence the Soret effect. 
Polymers are examples of
molecules that have large internal 
degrees of freedom \cite{
Giddings_Macro_1976,
Schimpf_JPCB_1989,
Zhang_JCP_2006,
Wuerger_PRL_2007,
Stadelmaier_Macro_2009,
Wuerger_PRL_2009}. 
If a polymer is much longer than its persistent length, 
it behaves as a flexible chain;
it behaves as a rigid rod if it is shorter 
than the persistent length. 
It was reported that 
the Soret coefficient of the polymer solutions
changes with the chain length 
when it is short. 
However, when it is sufficiently long, 
the Soret coefficient is saturated to a value 
that is independent of the chain length.
For the polymers, 
the mass and the moment of inertia change with 
the chain length. 
Although 
some theoretical studies  
on the Soret effect of the polymer systems 
have been reported, 
the pure contribution of the flexibility of the 
molecules to the Soret effect is still unclear. 
The aim of this work is to investigate the 
roles of the internal degrees of freedom 
in the Soret effect by employing the simplest 
molecular model. 

\section{Molecular dynamics simulation}

We perform molecular dynamics simulations of mixtures of 
two molecular species in three dimensional rectangular boxes 
($V=L_x L_y  L_z$). 
One species is the simplest molecule, which consists of a 
single spherical particle; 
the other is a dimer, which consists of two identical spherical 
particles. 
The particles in the monomers and dimers interact mutually 
with the Weeks-Chandler-Andersen (WCA) potential \cite{Weeks_JCP_1971}, 
which is given by 
\bea
U_{\rm WCA}(r)=
\left\{
\begin{array}{ll}
4\epsilon 
\left[(\sigma/r)^{2n}-(\sigma/r)^n+1/4\right]  & (r\le 2^{1/n}\sigma) \\
0 & (r>2^{1/n}\sigma)
\end{array}
\right., 
\nonumber\\
\ena
where $\epsilon(=1)$ and $\sigma(=1)$ represent the strength 
and range of the WCA potential. 
$n$ is a parameter for characterizing the hardness of the WCA potential. 
If we do not mention explicitly, we set $n=6$. 
The two particles in each dimer are bounded by a harmonic potential, 
\bea
U_{\rm sp}(r)=k(r-r_0)^2, 
\ena
where $k$ is the spring constant, 
while 
$r_0$ is the natural length of the bond, and 
we set $r_0=2^{1/n}\sigma$. 
The WCA potential between the particles in each dimer is not included. 
The mass of the particles is given by $m$; thus the mass of the 
dimer is $2m$.
The numbers of the monomer and dimer molecules are given by $N_{\rm m}$ 
and $N_{\rm d}$.
The particle packing fraction is 
defined by $\phi=\pi \sigma^3N_{\rm t}/(6V)$, where $N_{\rm t}$ is 
the total particle number, $N_{\rm t}=N_{\rm m}+2N_{\rm d}$. 
The mixing ratio of the dimer is given by $\chi=2N_{\rm d}/N_{\rm t}$.

A temperature gradient is imposed along 
the $x$-axis by using boundary driven non-equilibrium 
molecular dynamics simulation. 
The thermostatting regions are set up at the edges ($x=0$ and $x=L_x$) 
and at the center ($x=L_x/2$) of the rectangular cell, 
while their width is $0.03L_x$.
The temperatures in these thermostatting regions are imposed 
to $T=T_{\rm h}$ at the edges and $T=T_{\rm c}$ at the center 
($T_{\rm h}>T_{\rm c}$) by means of the Langevin thermostat. 
In the other bulk regions, the particle position and velocity are 
updated without any thermostat. 
The equations of motion are solved with velocity Verlet algorithm 
using LAMMPS \cite{LAMMPS}, in which 
the time increment is $\delta t=0.0005\sqrt{\epsilon/(m\sigma^2)}$. 
We fix the cell width as $L_x=80\sigma$, while $L_y$ and $L_z$ are 
changed to adjust the packing fraction $\phi$. 
The total particle number is $N_{\rm t}=64000$, and 
the packing fraction  
is changed from $\phi=0.037$ to $0.234$. 
If we do not mention explicitly, we set 
the mixing ratio to $\chi=1/2$.
With the SHAKE algorithm \cite{Rapaport_book_2013}, 
we also consider dimers, in which the particle separations are 
fixed to $r=r_0$. 
To observe the pure effect of the dimers, we also study 
the Soret effect in mixtures of two monomers A and B, among which the 
masses or the radii differ. 

\section{3D simulation results}

Figure~1(a) shows the profiles of the kinetic energies along 
the $x$-axis. 
The temperatures in the thermostatted regions are 
set to $T_{\rm h}=5\epsilon $ and $T_{\rm c}=\epsilon$. 
We divide the cell into 100 slabs along the $x$-axis, 
and then, 
we calculate the local densities of 
the monomer particle $\rho_{\rm m}(x)$ and the dimer 
particle $\rho_{\rm d}(x)$,  
and the kinetic energies averaged per particle 
in each thin slab.
Here, $\int {\rm d}x\rho_{\rm d}(x)=2N_{\rm d}$ 
and $\int {\rm d}x\rho_{\rm m}(x)=N_{\rm m}$ are held.
We plot the translational kinetic energies of 
the monomer $K_{\rm m}^{\rm tra}$, and 
the three modes (translational $K_{\rm d}^{\rm tra}$, 
rotational $K_{\rm d}^{\rm rot}$, 
and vibrational $K_{\rm d}^{\rm vib}$) 
of the kinetic energies of the dimer. 
In the equilibrium state, $K_{\rm m}^{\rm tra}$, 
$K_{\rm d}^{\rm tra}$, $K_{\rm d}^{\rm rot}$, and
$K_{\rm d}^{\rm vib}$ should agree with 
$3T/2$, $3T/2$, $T$, and $T/2$, respectively. 
It can be seen 
that all the averaged kinetic energies collapse on a master curve 
after appropriate scalings (see the inset of Fig.~1(a)). 
The kinetic energies in the thermostated regions are consistent with the 
target temperatures $T_{\rm h}$ and $T_{\rm c}$ 
and  vary linearly with $x$ in the bulk regions. 
Thus, the temperature is  controlled well in our system. 

\begin{figure}
\includegraphics[width=0.35\textwidth]{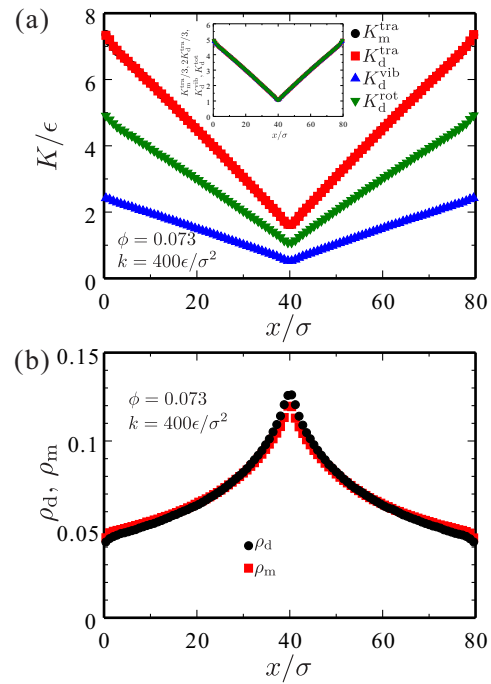}
\caption{
(a) Profiles of the local kinetic energies of the monomers ($K_{\rm m}^{\rm tra}$) and the dimers ($K_{\rm d}^{\rm tra},\, K_{\rm d}^{\rm vib}$, and $K_{\rm d}^{\rm rot}$) along the temperature gradient. 
The average packing fraction and the spring constant are $\phi=0.073$ and $k=400\epsilon/\sigma^2$, respectively. 
The temperatures at the thermostatted regions are $T_{\rm h}=5\epsilon$ 
and $T_{\rm c}=\epsilon$. 
In the inset, the scaled kinetic energies are replotted.
(b) Profiles of the local densities of the dimer $\rho_{\rm d}$ and the 
monomer $\rho_{\rm m}$. 
The parameters are the same as those in (a). 
}
\end{figure}

Figure~1(b) plots the local density profiles of the dimer 
$\rho_{\rm d}(x)$ and the monomer $\rho_{\rm m}(x)$. 
The spring constant is $k=400\epsilon/\sigma^2$, 
while the packing fraction is $\phi=0.073$. 
Both densities are higher in the colder 
than in the hotter regions. 
It is also indicated that the density 
of the dimers is slightly higher than that of the monomers 
in the cold region. 
The Soret effect is induced in this 
mixture of dimers and monomers. 

Figure~2(a) illustrates the profiles of the concentration field of the 
dimer in a dilute mixture, 
which is defined as $c(x)=\rho_{\rm d}(x)/(\rho_{\rm m}(x)+\rho_{\rm d}(x))$. 
If the concentration is homogeneous, its value agrees with the mixing 
ratio, {\it i.e.,} $c(x)=\chi$. 
The total packing fraction and the spring constant are set to 
$\phi=0.073$ 
and $k=400\epsilon/\sigma^2$, respectively. 
In Fig.~2(a), we change the temperature difference 
$\Delta T=T_{\rm h}-T_{\rm c}$ by fixing the average 
temperature to $\av{T}(=(T_{\rm h}+T_{\rm c})/2)=3\epsilon$. 
In the absence of the temperature difference ($\Delta T=0$), 
the concentration is almost constant in space (not shown here).
As  $\Delta T$ increases,
the concentration near the cold region increases, 
while that near the hot region decreases.
This 
means that the dimers tend to migrate to the colder side. 
The degree of dimer migration  is almost 
proportional to the temperature difference. 
Figure~2(b) presents the profiles of the concentration in 
a dense mixture ($\phi=0.234$). 
In contrast to the dilute mixture, the dimers migrate 
to the hotter region in the dense mixture. 
The degree of dimer migration is almost 
proportional to the temperature difference. 

\begin{figure}
\includegraphics[width=0.35\textwidth]{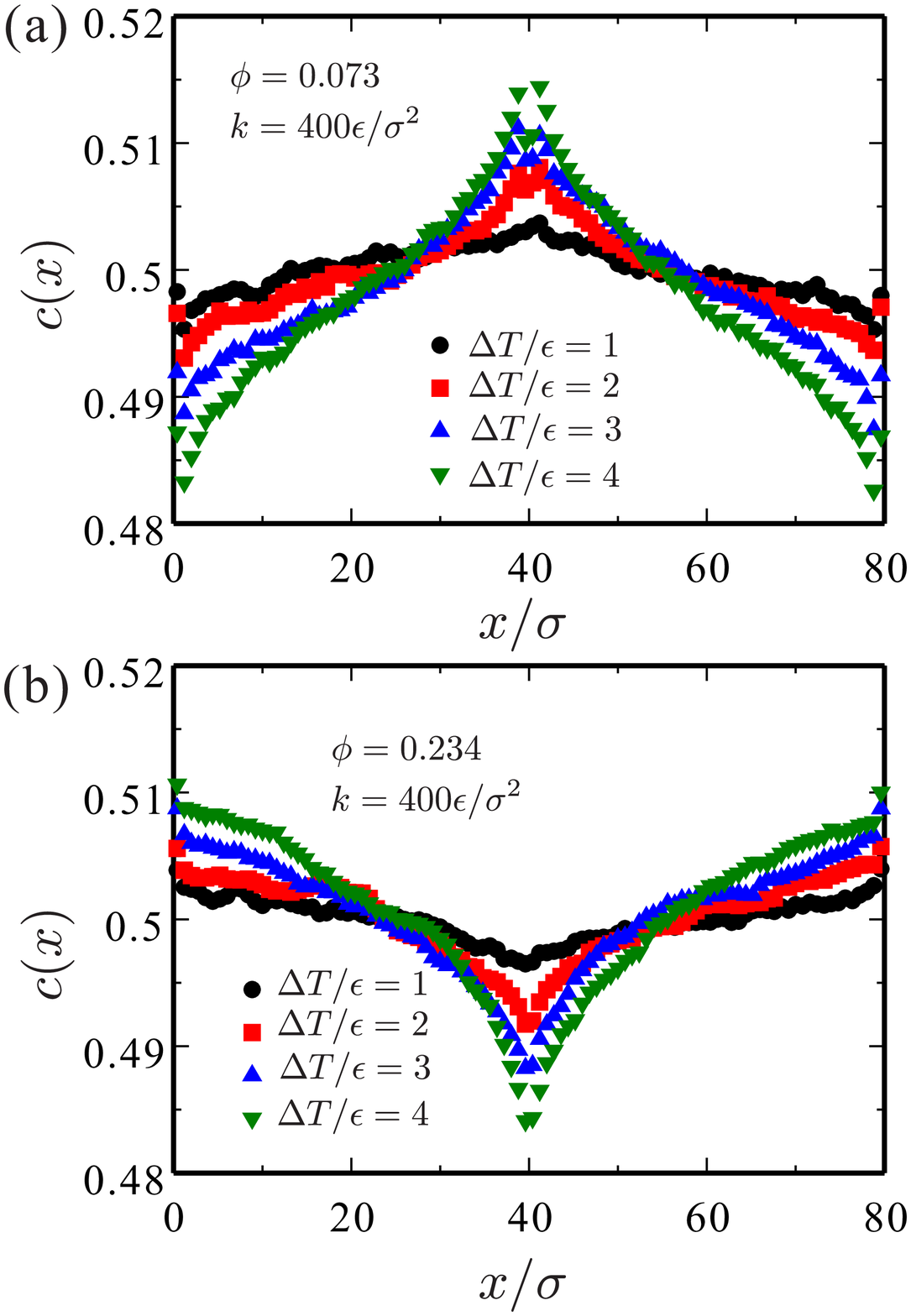}
\caption{
(a) Profiles of the concentration of the dimers $c=\rho_{\rm d}/(\rho_{\rm d}+\rho_{\rm m})$ in a dilute mixture of $\phi=0.073$. 
The temperature difference $\Delta T=T_{\rm h}-T_{\rm c}$ is changed. 
(b) Profiles of the concentration 
of the dimers in a dense mixture of 
$\phi=0.234$. 
}
\end{figure}

Instead of the Soret coefficient $S_T$, 
we analyze the Soret effect with 
the thermal diffusion factor $\alpha_T =TS_T$.
We evaluate it 
from the local thermal diffusion factor, 
\begin{eqnarray}
\alpha(x)=-\frac{T}{c(1-c)}
\frac{\partial c/\partial x}{\partial T/\partial x},
\end{eqnarray} 
and average $\alpha(x)$ over the system to get 
$\alpha_T=\int {\rm d}x\alpha(x)/L_x$. 
A positive value of $\alpha_T$ implies the dimer migrates to the 
cold side. 
Figure~3(a) shows the dependences of $\alpha_T$ against 
the packing fraction $\phi$.
The spring constant in the dimers is changed from 
$k=10\epsilon/\sigma^2$ to $500\epsilon/\sigma^2$. 
$\alpha_T$ for the rigid dimer is also given in the same figure. 
In Fig.~3(b), we replot $\alpha_T$ as functions of the spring constant $k$ 
for several densities.

In dilute systems, the thermal diffusion factors are positive 
for any $k$, which 
indicates that the dimers migrate to the colder side, 
as illustrated in Fig.~2(a). 
As the packing fraction $\phi$ 
increases, the thermal diffusion factors decrease. 
Here the decreasing rate of $\alpha_T$ against $\phi$
is larger 
in the mixture of smaller $k$ than in that of larger $k$. 
For smaller $k$, 
the thermal diffusion factor changes its sign 
from positive to negative with $\phi$. 
In contrast, the thermal diffusion factor for the rigid dimer 
remains positive 
in the simulated range of $\phi$. 
The molecule becomes {more rigid, 
and the thermal diffusion factor is likely to converge  for the 
rigid molecules  
when the spring constant is increased. 
In Fig.~3(b), the thermal diffusion factor changes its sign with the 
spring constant $k$ in the intermediated mixtures 
($\phi=0.146$). 
Our results clearly demonstrate that 
the internal degrees of freedom in the molecules can influence 
the Soret effect, which is usually considered 
to be an inter-molecular phenomenon.

\begin{figure}
\includegraphics[width=0.35\textwidth]{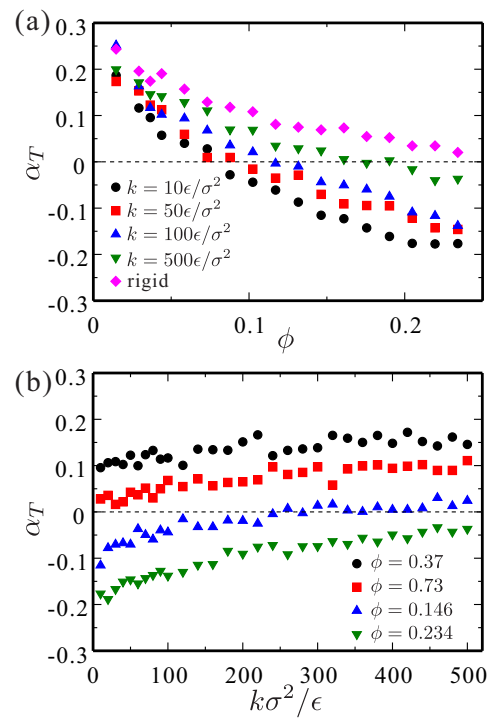}
\caption{
(a) Plots of the thermal diffusion factor $\alpha_T$ 
against the packing fraction  $\phi$. 
The spring constant is changed. 
$\alpha_T$ for the mixture of the rigid dimer, and monomer 
is also presented. 
(b) Plots of $\alpha_T$ with respect to the spring constant $k$. 
The packing fraction  is changed. 
}
\end{figure}

\section{Discussion}

First, we consider the mechanism of the Soret effect in terms of the isotope effect. 
In Fig.~4, we plot the thermal diffusion factor in mixtures of two monomers A and B (black circles). 
In these mixtures, 
the interactions among the particles are given by Eq.~(1), while 
the mass of the A monomer is twice larger than that of the B monomer, 
{\it i.e.,} $\sigma_{\rm AA}=\sigma_{\rm BB}=\sigma_{\rm AB}=\sigma$, 
with $m_{\rm A}=2m$ and $m_{\rm B}=m$. 
In Fig.~4, the positive values of the thermal diffusion factor agree with 
the isotope effect;
heavier particles A migrate to the colder region. 
In contrast to the decreasing thermal diffusion factor in the dimer-monomer 
mixtures (Fig.~3(a)), the thermal diffusion factor is increased with 
the packing fraction  in the monomer-monomer mixtures. 
Thus, it is concluded that the Soret effect in the dimer-monomer mixtures 
is essentially different from the isotope effect. 
It is also known that molecules of larger inertia moments migrate 
to the colder region \cite{Artola_JACS_2008,Debuschewitz_PRL_2001,Galliero_FPE_2003}. 
If the Soret effect observed in this study is due to the 
inertia effect, the thermal diffusion factor would be increased by using 
longer natural length of the bond interaction in the dimers (Eq.~(2)). 
However, we confirmed that the thermal 
diffusion factor is increased when the natural 
length in the bond interaction is shortened (not shown here). 
In particular, if we set $r_0=0$ in Eq.~(2) and use larger $k$, 
the thermal diffusion factor is close to that due to the 
isotope effect. 
Thus, we believe that 
the Soret effect in our mixtures is not due to the 
inertia effect.

\begin{figure}
\includegraphics[width=0.35\textwidth]{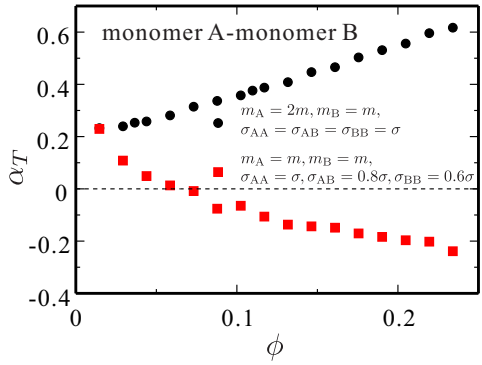}
\caption{
The thermal diffusion factors in the mixtures of two monomers A and B 
are plotted against $\phi$. 
black circle: mixtures of monomers with different masses; 
red square: mixtures of monomers with different radii. 
}
\end{figure}

Next we consider the contribution of the thermodynamic chemical potential 
to the Soret effect in our simulations. 
The linear non-equilibrium thermodynamic theory gives the particle flux $\vec{J}_i$ ($i= $d or m) as \cite{deGroot_book_1985}
\begin{eqnarray}
 \vec{J}_i=-\rho_iL_{1i}\nabla \frac{\tilde{\mu}_i}{T}+\rho_iL_{2i}  \nabla \frac{1}{T},
\end{eqnarray}
where $L_{1i}$ and $L_{2i}$ are phenomenological kinetic coefficients, 
and $\tilde{\mu}_i$ is an effective chemical potential.  
In the steady state, it should vanish, {\it i.e.,} $\vec{J}_i=0$. 
If $L_{2i}=0$, the Soret effect is determined by the thermodynamic 
properties of the chemical potential against $\chi$ and $T$. 
In Figs.~5(a), (b) and (c), we represent the effective chemical potential 
of the dimer $\tilde{\mu}_{\rm d}$ against  
the mixing ratio $\chi$, the temperature $T$, 
and the packing fraction $\phi$, 
respectively.  
In Fig.~5(a), we changed the mixing ratio $\chi$ with fixing the total number $N_{\rm t}$. 
The effective chemical potential is given as 
$\tilde{\mu}_{\rm d}=\mu_{\rm d}-(v_{\rm d}/v_{\rm m})\mu_{\rm m}$, where $\mu_i$ and $v_i$ are the chemical potential and partial molecular 
volume of the $i$-component. 
In Fig.~5, we assume $v_{\rm d}/v_{\rm m}=2$, for simplicity. 
The chemical potentials $\mu_{\rm d}$ and $\mu_{\rm m}$ are 
are obtained through the Widom insertion method \cite{Widom_JCP_1963} 
in the simulations 
without the temperature difference. 
As indicated in Fig.~5(a), the chemical potential difference is almost 
independent of the mixing ratio $\chi$, 
while $\tilde{\mu}_{\rm d}/T$ is an increasing function of $T$ in Fig.~5(b). 
The chemical potential modulation induced by the temperature gradient cannot 
be compensated by the chemical potential change with the mixing ratio. 
This means that 
the Soret effect in our system is not caused by the 
thermodynamic chemical potentials, 
and it is purely a non-equilibrium behavior. 
The chemical potential difference also depends on the  packing fraction 
in Fig.~5(c). 
$\partial (\tilde \mu_{\rm d}/T)/\partial T$ is compensated roughly by 
$\partial (\tilde \mu_{\rm d}/T)/\partial \phi$ (see Fig.~1(b)).

\begin{figure}
\includegraphics[width=0.35\textwidth]{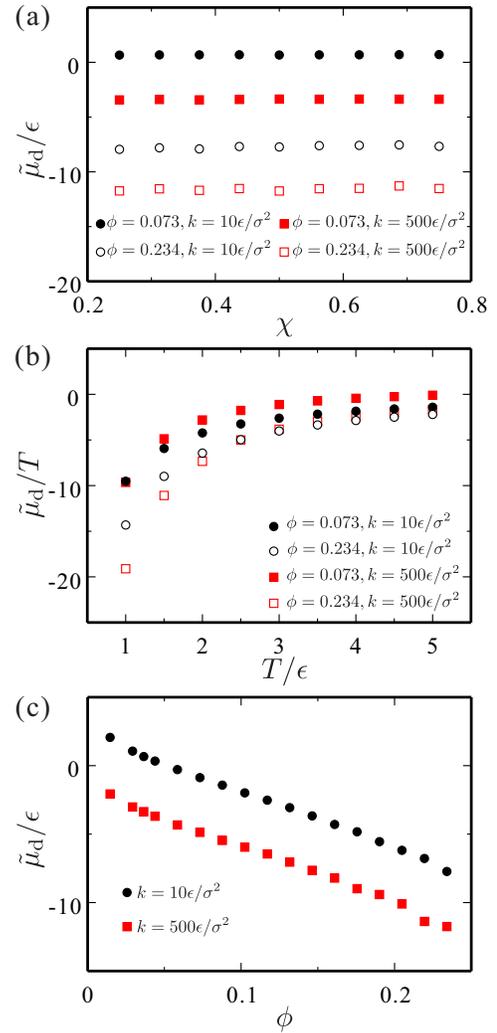}
\caption{
The differences of the chemical potentials between the 
dimers and monomers $\Delta \mu$ are plotted 
against the temperature $\av{T}$, 
the mixing ratio $\chi$,
and the packing fraction  $\phi$ in 
(a), (b) and (c), respectively. 
In (a) and (c), the temperature is $T=3\epsilon$.}
\end{figure}

As discussed above, we cannot understand our simulation results with 
conventional knowledge of the Soret effect, {\it i.e.}, 
the contributions of the mass, inertia moment and chemical potential.  
Here, we propose possible mechanisms of 
the Soret effect in our dimer-monomer mixtures. 
In Fig.~6(a), 
we present the probability distribution $P(r)$ of the bond length $r$ of the dimers 
in the mixtures of $\phi=0.036$ and $0.234$. 
The spring constant is $k=10\epsilon/\sigma^2$ and the 
temperature is changed. 
When we obtain $P(r)$, the temperature gradients are not imposed. 
As the temperature is increased, not only the distribution width 
is broadened, but also the peak position is shifted to large $r$. 
We plot their average lengths $\av{r}$ against the temperature in 
Fig.~6(b). 
Those for $k=500\epsilon/\sigma^2$ are also plotted. 
The dimer molecule is more stretched at the high temperature. 
In other words, the dimer becomes bulkier. 
The degree of the stretching is large when $k$ is small, so that 
the corresponding effective volume of the dimer is increased more 
largely for small $k$. 
In the dilute mixtures ($\phi=0.036$), the probability 
distribution agrees well with 
the Boltzmann distribution ($4\pi r^2  \times\exp[-U_{\rm sp}(r)/T]$). 
In the dense mixtures, 
on the other hand, the probability distribution is inconsistent with 
the Boltzmann distribution. 
The average bond length in the dense mixture is shorter than 
that in the dilute mixture as shown in Fig.~6(b). 
The molecules are surrounded by other molecules, all of which are 
exerting non-bonded forces of the molecules. 
The degree of the bond stretching is suppressed more in the dense system.

In Fig.~4, we also plotted the thermal 
diffusion factor in other monomer-monomer mixtures 
as a function of the packing fraction $\phi$ (red squares). 
In these mixtures, the masses of both monomers are the same, 
while the size of the A-particle is set to be larger than that of B-particle.  
We replace $\sigma$ in the WCA potential (Eq.~(1)) to 
$\sigma_{\rm AA}=\sigma,\,\sigma_{\rm AB}=0.8\sigma$, and 
$\sigma_{\rm BB}=0.6\sigma$ for the A-A, A-B and B-B pairs of the monomers, 
respectively.  
The potential strength $\epsilon$ is not changed. 
As the packing fraction  increases, the thermal diffusion factor decreases 
and becomes negative eventually. 
The bulky A-monomers migrate to the colder region in the dilute mixtures, 
while they move to the hotter region in the dense mixtures. 
These observations are consistent 
with a previous study \cite{Galliero_FPE_2003}. 
Although the physical mechanism of this Soret effect still remains unclear, 
we consider that the Soret effect in the dimer-monomer mixtures is related 
to that in these monomer-monomer mixtures. 
When $k$ is small, the dimer molecules become bulkier. 
Then the bulky dimers tend to move to the hotter region as that in the 
monomer-monomer mixtures, although these dimers are two times 
as heavy as the monomers.
If $k$ is large, the dimer size does not change much, 
making this behavior unremarkable.

\section{1D simulation}

In this section,  
we propose another possible mechanism of the Soret effect, 
with one-dimensional (1D) 
simulations \cite{Garriga_JSP_2002}. 
The motions of the two particles in a dimer and a monomer particle are 
constrained in the $x$-axis. 
We prepared initial dimer and monomer, which obey the Maxwell-Boltzmann 
distributions at $T=T_{\rm d}$ and $T=T_{\rm m}$, respectively.
The 1D dimer has three types of energies ($K_{\rm d}^{\rm tra}$, 
$K_{\rm d}^{\rm vib}$, 
and spring potential $U_{\rm sp}$), 
the statistical averages of which are set to $T_{\rm d}/2$. 
The translational kinetic energy of the monomer is set to 
$\av{K_{\rm m}^{\rm tra}}=T_{\rm m}/2$. 
The monomer particle and one dimer particle, 
which we call as the first particle (and the other as 
the second one), interact via 
the WCA potential (Eq.~(1)) with $n=6$. 
If the dimer and the monomer collide and bounce, 
we calculate the coefficient of restitution $e$ from 
the velocities of the dimer and monomers. 
The collision between the dimer and the monomer is inelastic, 
since the sum of the translational kinetic energies, 
$K_{\rm d}^{\rm tra}+K_{\rm m}^{\rm tra}$, 
is not conserved  after 
the collision.  
In particular, 
when the energy due to the internal degrees of freedom 
($K_{\rm d}^{\rm vib}+U_{\rm sp}$)
is transferred to the translational 
kinetic energies ($K_{\rm d}^{\rm tra}+K_{\rm m}^{\rm tra})$, 
the coefficient of restitution is likely to 
exceed unity \cite{Kuninaka_PRE_2012}. 
The coefficient of restitution is scattered statistically, 
so that we obtain its average with $10^7$ samples. 

\begin{figure}
\includegraphics[width=0.35\textwidth]{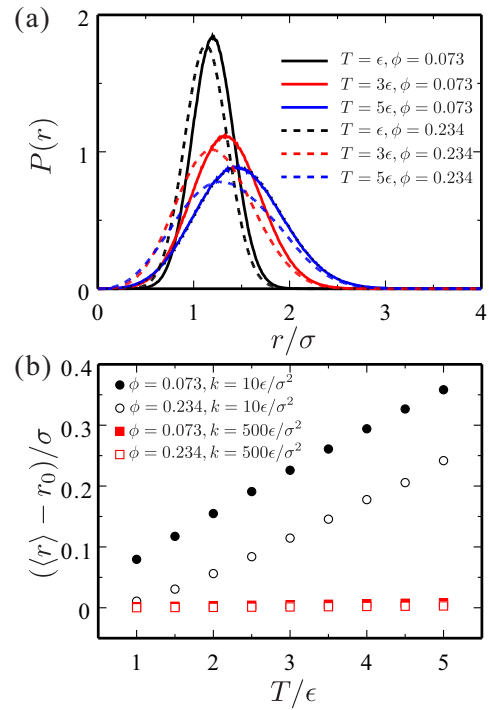}
\caption{
(a) 
Probability distributions $P(r)$ of the bond length $r$ 
in the dimer-monomer 
mixtures of $\phi=0.036$ and $0.234$. 
The spring constant is $k=10\epsilon/\sigma^2$. 
The temperature $T$ is changed. 
(b) Plots of the averaged bond length $\av{r}$ with respect to 
the temperature $\Delta T$ in the 
dimer-monomer mixtures. 
}
\end{figure}

In Fig.~7(a), we present 
the coefficients of restitution with respect to the spring constant $k$ 
in two cases, 
X: $(T_{\rm d}, T_{\rm m})=(5\epsilon, \epsilon)$, 
and Y: $(T_{\rm d}, T_{\rm m})=(\epsilon, 5\epsilon)$. 
In both cases, 
the coefficient of restitution is larger than unity when $k$ is small, 
and it approaches to unity when $k$ is large. 
It can be observed that the dimer and monomer bounce asymmetrically, 
{\it i.e.}, 
$e_{\rm X}>e_{\rm Y}$ over the entire range of $k$. 
In case X, the coefficient of restitution is decreased monotonically to unity 
as the spring constant is increased. 
In case Y, on the other hand, the coefficient of restitution changes non-monotonically with $k$. 
When $k$ is small, $e_{\rm Y}$ is decreased with $k$ 
and becomes smaller than unity. 
Then, it turns to increase and approaches unity for large $k$. 
In case X, the dimer is likely to have larger internal energy 
than the sum of the translational kinetic energies 
($K_{\rm d}^{\rm vib}+U_{\rm sp}>K_{\rm d}^{\rm tra}+K_{\rm m}^{\rm tra}$) before the collision.
Upon the collision, the internal energy changes to the translational energy, 
increasing the coefficient of restitution. 
In case Y, on the other hand, the internal energy is comparably small 
($K_{\rm d}^{\rm vib}+U_{\rm sp}<K_{\rm d}^{\rm tra}+K_{\rm m}^{\rm tra}$),
allowing the translational energy transferred to the internal energy. 
As a result, the coefficient of restitution in case Y is likely to be 
smaller than that in case X, $e_{\rm X}>e_{\rm Y}$. 
When the dimer is ``hotter" than the monomer, 
the molecules bounce more, resulting 
in the negative thermal diffusion factor, with which the dimers migrate 
to the hotter region.

In Fig.~7(b), we plot the difference between the coefficients of restitution 
in two different cases, $\Delta e=e_{\rm X}-e_{\rm Y}$.  
Here, we show $\Delta e$ 
for several different values of $n(=3,6,12$, and $24)$ 
in the WCA potential (Eq.~(1)). 
In any $n$, $\Delta e$ is positive and it decays with $k$. 
With small $n$, the particles are soft, so that the collision 
time increase, and the monomer and the first particle in the dimer 
interact via the WCA potential slowly. 
If the intra-particle interaction is strong enough (with large $k$), 
the energy and the momentum are well transferred 
to the second particle in the dimer during the collision. 
Meanwhile, the collision becomes more elastic when $n$ is small and $k$ 
is large, making 
the coefficients of restitution $e_{\rm X}$ and $e_{\rm Y}$ 
converge to unity and 
$\Delta e$ goes to zero. 
When $n$ is large, on the other hand, the collision time is shortened. 
If the intra-molecule interaction is weak (with small $k$), 
the second particle is negligible upon the collision. 
Since we set the particle masses to be same, 
the kinetic energies and momenta are exchanged between the 
monomer and the first particle  during the collision. 
However, after the collision, 
the translational and vibrational kinetic energies are redistributed within 
the dimer. 
If $T_{\rm d}=0$ in case Y, in particular, 
the coefficient of restitution is $e_{\rm Y}=1/2$ in this limit of large $n$. 
As a result, the difference between $e_{\rm X}$ and $e_{\rm Y}$ tends to zero 
more gradually when $n$ is large. 

\begin{figure}[h]
\includegraphics[width=0.35\textwidth]{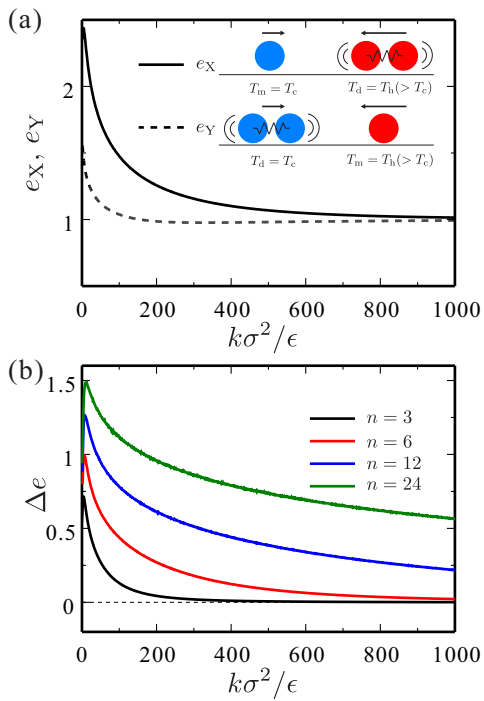}
\caption{
(a) 
The coefficients of restitution between a dimer and a monomer 
are shown against the spring constant $k$, 
which are obtained in one-dimensional simulations for 
two cases. 
X: $(T_{\rm d},T_{\rm m})=(5\epsilon,\epsilon)$, 
Y: $(T_{\rm d},T_{\rm m})=(\epsilon,5\epsilon)$.  
(b) The differences of the coefficients of restitution 
$\Delta e=e_{\rm X}-e_{\rm Y}$ is shown with the spring constant. 
The exponent $n$ in the WCA potential is changed. }
\end{figure}

Although we consider that this asymmetry of the restitutions can cause 
the Soret effect also in the actual systems, 
the relationship between the Soret effect in the 1D simulations and that in 
the above 3D simulations should be considered carefully. 
The collisions among the molecules 
in the 3D system are mostly oblique and not head-on. 
Thus, it is considered that 
the asymmetric restitution plays 
more minor roles in the three dimensional mixtures, 
even if it is the case. 
Because of the asymmetric restitution and 
the energy transfer, 
the thermalization of the internal and kinetic 
degrees of freedom might be violated. 
As shown in Fig.~1(a), however, we have not observed 
any relevant difference between 
the kinetic and vibrational temperatures. 
We consider that the linear temperature gradient are 
formed well after the particle heterogeneity 
due to the Soret effect is induced, 
although we have no evidence supporting this claim.

Here we mention a previous study of 
Garriga {\it et al.}\cite{Garriga_JSP_2002}, 
in which the authors have 
considered the Soret effect in the 1D 
binary mixtures of light and heavy particles. 
They found that the asymmetric collisions lead to 
the strong Soret effect. 
Although their molecular description is quite different 
from ours, 
their findings would support our conjecture.

\section{Summary}

In this article, 
we numerically studied the Soret effect in the dimer-monomer 
mixtures and 
found that the internal degrees of the freedom of 
the molecules can contribute significantly the Soret effect, 
which is usually treated as an inter-molecular phenomenon. 
In dense system, in particular, 
the rigid dimers migrate to the colder side, 
while the flexible ones move to the hotter side. 

To explain the Soret effect in our system, 
we proposed two possible mechanisms. 
One regards to the changes of the molecular shape; 
flexible molecules adapt their volume depending on the local 
temperature, 
while bulkier dimer molecules tend to migrate to the hotter side, 
overcoming the isotope effect. 
The other regards the asymmetry of the restitutions. 
If a molecule having larger internal degrees of freedom 
is ``hotter", 
the coefficient of restitution with the other ``colder" molecule 
with smaller internal degrees of freedom becomes larger than 
that for the opposite case. 
To determine which of them (and/or some other mechanisms) 
is more dominant, we have to investigate them further through 
other types of test molecules. 
The inter-molecular interactions would be modified by the 
intra-structures of the molecules in the non-equilibrium conditions.

In this work, we focused on dimer molecules consisting of two 
particles. 
However, it should be noted that 
diatomic molecules, such 
as ${\rm O}_2$, are not considered with this dimer model. 
The vibrational modes in a real diatomic molecule 
are quantised and the higher modes are strongly 
quenched at room temperature.
In this work, we aimed to consider softer molecules;
polymers are appropriate candidates for such soft molecules. 
However, the changes of the degree of the polymerization accompany 
those of the mass, the moment of inertia, and the interactions, 
all of which influence the Soret effect. 
Thus, it is difficult to observe the pure effect of the flexibility on 
the Soret effect. 
It was numerically demonstrated that the difference of the 
chain stiffness triggers the change of the Soret effect 
\cite{Zhang_JCP_2006}. 
When the degree of the polymerization is the same, 
the Soret coefficient of more flexible polymers, with short 
persistent length, tends to be smaller than 
those of rigid polymers with long persistent length. 
We believe that our findings can help us to further understand 
such mysterious behaviors of the Soret effect.

\section*{Acknowledgements}

This work was supported by KAKENHI (Grants No.17K05612), 
CREST, JST (JPMJCR1424, JPMJCR2095).


\begin{thebibliography}{99}

\bibitem{deGroot_book_1985}
S. R. de Groot and P. Mazur, {\it Non-equilibrium Thermo-
dynamics} (Dover Publications Inc., 1985).


\bibitem{Wiegand_JPCM_2004}
S. Wiegand, J. Phys.: Condens. Matter {\bf 16}, R357 (2004).

\bibitem{Platten_JAM_2005}
J. K. Platten, J. Appl. Mech. {\bf 73}, 5 (2005).

\bibitem{Srinivasan_book_2012}
S. Srinivasan and M. Z. Saghir, {\it Thermodiffusion in Mul-
ticomponent Mixtures} (Springer New York, 2012).


\bibitem{Kita_JCP_2004}
R. Kita, S. Wiegand, and J. Luettmer-Strathmann, J. Chem. Phys. {\bf 121}, 3874 (2004).

\bibitem{Perronace_JCP_2002}
A. Perronace, C. Leppla, F. Leroy, B. Rousseau, and S. Wiegand, J. Chem. Phys. {\bf 116}, 3718 (2002).


\bibitem{Giddings_Macro_1976}
J. C. Giddings, K. D. Caldwell, and M. N. Myers, Macromolecules {\bf 9}, 106 (1976).

\bibitem{Schimpf_JPCB_1989}
M. E. Schimpf and J. C. Giddings, J. Polym. Sci. B Polym. Phys. {\bf 27}, 1317 (1989).


\bibitem{Zhang_JCP_2006}
M. Zhang and F. M\"{u}ller-Plathe, J. Chem. Phys. {\bf 125}, 124903 (2006).



\bibitem{Wuerger_PRL_2007}
A. W\"{u}rger, Phys. Rev. Lett. {\bf 98}, 138301 (2007).

\bibitem{Stadelmaier_Macro_2009}
D. Stadelmaier and W. K\"{o}hler, Macromolecules {\bf 42}, 9147 (2009).

\bibitem{Wuerger_PRL_2009}
A. W\"{u}rger, Phys. Rev. Lett. {\bf 102}, 078302 (2009).

\bibitem{Piazza_SM_2008}
R. Piazza, Soft Matter {\bf 4}, 1740 (2008)

\bibitem{Duhr_PNAS_2006}
S. Duhr and D. Braun, Proc. Natl. Acad. Sci. U. S. A. {\bf 103}, 19678 (2006).

\bibitem{Wienken_NatComm_2010}
C. J. Wienken, P. Baaske, U. Rothbauer, D. Braun, and S. Duhr, Nat. Commun. {\bf 1}, 100 (2010).


\bibitem{Eastman_JACS_1928}
E. D. Eastman, J. Am. Chem. Soc. {\bf 50}, 283 (1928).

\bibitem{Artola_JACS_2008}
P.-A. Artola, B. Rousseau, and G. Galli\'{e}ro, J. Am. Chem. Soc. {\bf 130}, 10963 (2008).

\bibitem{Artola_MP_2013}
P.-A. Artola and B. Rousseau, Mol. Phys. {\bf 111}, 3394 (2013).

\bibitem{Hafskjold_MP_1993}
B. Hafskjold, T. Ikeshoji, and S. K. Ratkje, Mol. Phys. {\bf 80}, 1389 (1993).

\bibitem{Kempers_JCP_1989}
L. J. T. M. Kempers, J. Chem. Phys. {\bf 90}, 6541 (1989).

\bibitem{Garriga_JSP_2002}
A. Garriga, J. Kurchan, and F. Ritort, J. Stat. Phys. {\bf 106}, 109 (2002).

\bibitem{Wuerger_CRS_2013}
A.W\"{u}rger, Comptes Rendus M\'{e}canique {\bf 341}, 438 (2013).

\bibitem{Galliero_FPE_2003}
G. Galli{\'{e}}ro, B. Duguay, J.-P. Caltagirone, and F. Montel, Fluid Ph. Equilibria {\bf 208}, 171 (2003). 

\bibitem{Debuschewitz_PRL_2001}
C. Debuschewitz and W. K\"{o}hler, Phys. Rev. Lett. {\bf 87}, 055901 (2001).

\bibitem{Reith_JCP_2000}
D. Reith and F. M\"{u}ller-Plathe, J. Chem. Phys. {\bf 112}, 2436 (2000).


\bibitem{VillainGuillot_PRE_2011}
S. Villain-Guillot and A. W\"{u}rger, Phys. Rev. E {\bf 83}, 030501(R) (2011).

\bibitem{Weeks_JCP_1971}
J. D. Weeks, D. Chandler, and H. C. Andersen, J. Chem. Phys. {\bf 54}, 5237 (1971).

\bibitem{LAMMPS}
LAMMPS molecular dynamics package WWW site: lammps.sandia.gov.

\bibitem{Rapaport_book_2013}
D. Rapaport, {\it The Art of Molecular Dynamics Simulation} (Cambridge University Press, 2013).

\bibitem{Widom_JCP_1963}
B. Widom, J. Chem. Phys. {\bf 39}, 2808 (1963).

\bibitem{Kuninaka_PRE_2012}
H. Kuninaka and H. Hayakawa, Phys. Rev. E {\bf 86}, 051302 (2012).


\end{thebibliography}
\end{document}